\documentclass{elsart}
\usepackage{epsfig}
\usepackage{amssymb}
\usepackage{rotating}

\begin{document}

\begin{frontmatter}

\title{The ZEPLIN-III dark matter detector:\\
  performance study using an end-to-end simulation tool}

\author[ICL,RAL]{H. M. Ara\'ujo\corauthref{cor1}},
\corauth[cor1]{Corresponding author; address: Astrophysics Group,
Blackett Laboratory, Imperial College London, SW7 2BW, UK}
\ead{H.Araujo@imperial.ac.uk}
\author[ITP]{D. Yu. Akimov},
\author[RAL]{G. J. Alner},
\author[ICL]{A. Bewick},
\author[RAL]{C. Bungau},
\author[RAL]{B. Camanzi},
\author[SHE]{M. J. Carson},
\author[LIP]{V. Chepel},
\author[SHE]{H. Chagani},
\author[ICL]{D. Davidge},
\author[SHE]{J. C. Davies},
\author[SHE]{E. Daw},
\author[ICL]{J. Dawson},
\author[RAL]{T. Durkin},
\author[ICL,RAL]{B. Edwards},
\author[SHE]{T. Gamble},
\author[EDI]{C. Ghag},
\author[SHE]{R. Hollingworth},
\author[ICL]{A. S. Howard},
\author[ICL]{W. G. Jones},
\author[ICL]{M. Joshi},
\author[SHE]{J. Kirkpatrick},
\author[ITP]{A. Kovalenko},
\author[SHE]{V. A. Kudryavtsev},
\author[ICL]{V. N. Lebedenko},
\author[SHE]{T. Lawson},
\author[RAL]{J. D. Lewin},
\author[SHE]{P. Lightfoot},
\author[LIP]{A. Lindote},
\author[ICL]{I. Liubarsky},
\author[LIP]{M. I. Lopes},
\author[RAL]{R. L\"{u}scher},
\author[SHE]{P. Majewski},
\author[SHE]{K. Mavrokoridis},
\author[SHE]{J. McMillan},
\author[SHE]{B. Morgan},
\author[SHE]{D. Muna},
\author[EDI]{A. S. Murphy},
\author[LIP]{F. Neves},
\author[SHE]{G. Nicklin},
\author[SHE]{S. Paling},
\author[LIP]{J. Pinto da Cunha},
\author[RAL]{R. Preece},
\author[ICL]{J. J. Quenby},
\author[SHE]{M. Robinson},
\author[LIP]{C. Silva},
\author[LIP]{V. N. Solovov},
\author[RAL]{N. J. T. Smith},
\author[RAL]{P. F. Smith},
\author[SHE]{N. J. C. Spooner},
\author[ITP]{V. Stekhanov},
\author[ICL]{T. J. Sumner},
\author[SHE]{D. R. Tovey},
\author[ICL]{C. Thorne},
\author[SHE]{E. Tziaferi} \&
\author[ICL]{R. J. Walker}

\address[ICL] {Blackett Laboratory, Imperial College London, UK}
\address[RAL] {Particle Physics Department, Rutherford Appleton Laboratory, Chilton, UK}
\address[ITP] {Institute for Theoretical and Experimental Physics, Moscow, Russia}
\address[LIP] {LIP--Coimbra \& Department of Physics, University of Coimbra, Portugal}
\address[SHE] {Department of Physics \& Astronomy, University of Sheffield, UK}
\address[EDI] {Department of Physics \& Astronomy, University of Edinburgh, UK}

\newpage

\begin{abstract}
We present results from a GEANT4-based Monte Carlo tool for end-to-end
simulations of the ZEPLIN-III dark matter experiment. ZEPLIN-III is a
two-phase detector which measures both the scintillation light and the
ionisation charge generated in liquid xenon by interacting particles
and radiation. The software models the instrument response to
radioactive backgrounds and calibration sources, including the
generation, ray-tracing and detection of the primary and secondary
scintillations in liquid and gaseous xenon, and subsequent processing
by data acquisition electronics. A flexible user interface allows easy
modification of detector parameters at run time. Realistic datasets
can be produced to help with data analysis, an example of which is the
position reconstruction algorithm developed from simulated data. We
present a range of simulation results confirming the original design
sensitivity of a few times $10^{-8}$~pb to the WIMP-nucleon
cross-section.
\end{abstract}


\end{frontmatter}

% Main text %%%%%%%%%%%%%%%%%%%%%%%%%%%%%%%%%%%%%%%%%%%%%%%%%%%%%%%%%%%

%%%%%%%%%%%%%%%%%%%%%%%%%%%%%%%%%%%%%%%%%%%%%%%%%%%%%%%%%%%%%%%%%%%%%%%
\section{Introduction}

ZEPLIN-III is a two-phase (liquid/gas) xenon detector developed by the
UK Dark Matter Collaboration and international
partners,\footnote{Edinburgh University, Imperial College London,
ITEP-Moscow, LIP-Coimbra, Rutherford Appleton Laboratory and Sheffield
University.} which will try to identify and measure galactic dark
matter in the form of Weakly Interacting Massive Particles, or WIMPs
\cite{sumner05,araujo05a}. Upon completion of physics testing now
underway at Imperial College London, the system may join the ZEPLIN-II
\cite{zeplin2} and DRIFT-IIa \cite{drift2} experiments already
operating 1100~m underground in our laboratory at the Boulby mine
(North Yorkshire, UK).

Two-phase emission detectors based on the noble gases date back
several decades \cite{dolgoshein70}, but this technology has gained
momentum since the ZEPLIN Collaboration first explored the potential
of high-field xenon systems \cite{sumner99,howard01,akimov03}. The
operating principle is that different particle species generate
different amounts of scintillation light and ionisation charge in
liquid xenon (LXe). The ratio between these two signal channels
provides a powerful technique to discriminate between electron and
nuclear recoil interactions. WIMPs are expected to scatter elastically
off Xe atoms, much like neutrons, and the recoiling nucleus will
produce a different signature to $\gamma$-ray interactions and other
sources of electron recoils.

WIMP detectors differ from more traditional detectors of nuclear
radiation in that they require: i) extremely low radioactive and
cosmic-ray backgrounds, addressed by the use of radio-pure materials
and operation deep underground; ii) excellent discrimination of the
remaining background events, hopefully better than 1000:1 rejection
efficiency for electron recoils; iii) a low energy threshold for
nuclear recoils, since the kinematics of WIMP-nucleus scattering
results in a very soft recoil spectrum ($\lesssim$100~keV).

Monte Carlo simulations are essential in these key areas.  Acceptable
levels of trace contamination must be assessed for all detector
materials, requiring simulations of internal and external backgrounds
expected from each component. Cosmic-ray-induced backgrounds also need
careful calculation, since experimental measurements would require
nothing short of a dedicated WIMP detector. Having established the
residual electron/photon and neutron event rates, the level of
discrimination and energy threshold which can realistically be
achieved must be calculated with the help of detailed detector
simulations -- and possibly fed back to the design process. In
addition, the data produced by two-phase detectors are often complex,
and particular simulations are required to help extract actual physics
parameters. Finally, realistic datasets help with planning the data
acquisition electronics and the data analysis software.

In this paper we describe a simulation tool used throughout this
process \cite{zepIII}, based on the GEANT4 Monte Carlo toolkit
\cite{geant4}. The package builds upon previous simulation work
\cite{davidge03,dawson03,dmx} and experimental measurements with
high-field, two-phase prototype detectors
\cite{howard01,akimov03}. After overviewing the detector and the
software, a description is given of the simulation models used to
calculate the detector response. Simulation results are then presented
for the dominant background contributions, calibration runs with
$\gamma$ and neutron sources, and position reconstruction
capabilities, leading to a predicted performance of the instrument as
a WIMP detector.

%%%%%%%%%%%%%%%%%%%%%%%%%%%%%%%%%%%%%%%%%%%%%%%%%%%%%%%%%%%%%%%%%%%%%%%%%%%%%%
\subsection{The ZEPLIN-III detector}

The WIMP target is housed in a 1-m tall vacuum cryostat, which
contains a xenon vessel sitting on top of a liquid nitrogen
reservoir. The latter cools the xenon chamber to around
$-100{\rm^oC}$. All major metal components of the detector are made
from high-purity C101 copper. Inside the xenon vessel an array of 31
2-inch photomultiplier tubes (PMTs) is immersed in the liquid phase,
looking up to a $\simeq$40-mm thick LXe disk topped by a 5-mm layer of
gas. Up to 40~kV can be applied between a polished copper plate
covering the gas phase (hereafter `anode mirror') and a metal wire
grid located 35~mm below the liquid surface (`cathode grid'), defining
the active region of the detector. A second wire grid (`PMT grid')
placed just above the array (5~mm below the cathode) defines a reverse
field region which suppresses secondary signals from low-energy
photons from the PMTs and helps protect the PMT photocathodes from
stray electric fields. The PMTs, arranged in a closely-packed
hexagonal array, are encased in 2-inch holes bored in a copper screen
in order to prevent optical cross-talk. A lower plate (`PMT mirror')
covers the array; this mirror has conical cuts around the PMT windows
intended to improve the light collection and prevent the escape of
stray light generated around the PMT bodies. The diameter of the
active LXe volume is approximately 40~cm, while that of the PMT array
is 34~cm. The array is powered by a network of thin copper plates
located inside the xenon vessel, which provides common connection to
each dynode on all the PMTs and hence reduces the number of required
feedthroughs quite considerably. A detailed description of the
detector construction is given elsewhere \cite{sumner06}. We refer to
Fig.~\ref{geometry} for the model representation of the ZEPLIN-III
geometry.

When a particle interacts in LXe, VUV scintillation light is promptly
produced.  In addition, a strong electric field (up to
$\sim\!8$~kV/cm) prevents most of the ionisation charge produced
around the particle track from recombining. The extracted charge
carriers drift upward to the liquid surface, and are emitted into the
gas phase.  Here, the strong electric field (double that of the liquid
phase) leads to the production of further VUV photons, by the process
known as electroluminescence or proportional scintillation.
Therefore, both scintillation and ionisation signals give rise to
optical signatures which are detected with the same PMT array, the
time separation between them being proportional to the vertical ($z$)
coordinate. We label these signals as the `primary' and the
`secondary', or S1 and S2, respectively. The ratio S2/S1 is higher for
electrons than nuclear recoils. In low-energy atom-atom collisions,
most of the energy loss goes into recoil of the nuclei and only a
small fraction converts into electronic loss (i.e. electron
excitation/ionisation).  Electron/$\gamma$-ray interactions act
directly on the outer atomic electrons, yielding more ionisation
charge.

The 31 PMT signals are fed into wideband amplifiers and split into a
dual-range data acquisition system (DAQ). A large dynamic range
ensures sensitivity to very small primaries containing only a few
photoelectrons (phe) as well as large secondaries without
saturation. All 62 channels are sampled at 500~MS/s by 8-bit
digitisers.

%%%%%%%%%%%%%%%%%%%%%%%%%%%%%%%%%%%%%%%%%%%%%%%%%%%%%%%%%%%%%%%%%%%%%%%%%%%%%%
\subsection{The modelling software design}

The main requirements underlying the software design were: i) that it
should model the transport and interactions of particles internal and
external to the detector, down to the production of electron and
nuclear recoils; ii) simulate the physical processes involved in the
generation of the optical response to scintillation and
electroluminescence; iii) generate the electrical response (digitised
voltage timelines) for all channels in order to produce realistic
datasets; iv) operating parameters should be easily modified by the
user, in particular the LXe height and the applied electric field; v)
it should be user friendly.

GEANT4 is arguably the most comprehensive toolkit of its kind, and the
only one to fulfil the above requirements. Besides an extensive list
of physics models, its modular and transparent design means that new
ones can be easily added (e.g. electroluminescence). Its flexibility
allows further processing of simulated events (DAQ, event display,
data analysis, etc.). Versatile specification of the primary particle
generator means that individual particles, radioisotopes and sources
with complex spatial and energy distributions can be specified at run
time. Finally, user-interfaces are easily created, allowing the user
to control most physics parameters without having to delve into the
code itself. This is important since many optical and charge transport
properties are still ill-defined, and different parameter combinations
have to be assessed. In this package, primary generator, detector and
physics parameters can be set interactively by commands and macros
available for many simulation tasks.

%%%%%%%%%%%%%%%%%%%%%%%%%%%%%%%%%%%%%%%%%%%%%%%%%%%%%%%%%%%%%%%%%%%%%%%%%%%%%%
\section{The simulation model}

\subsection{GEANT4 solid model}

\begin{figure}[ht]
  \centerline{\epsfig{file=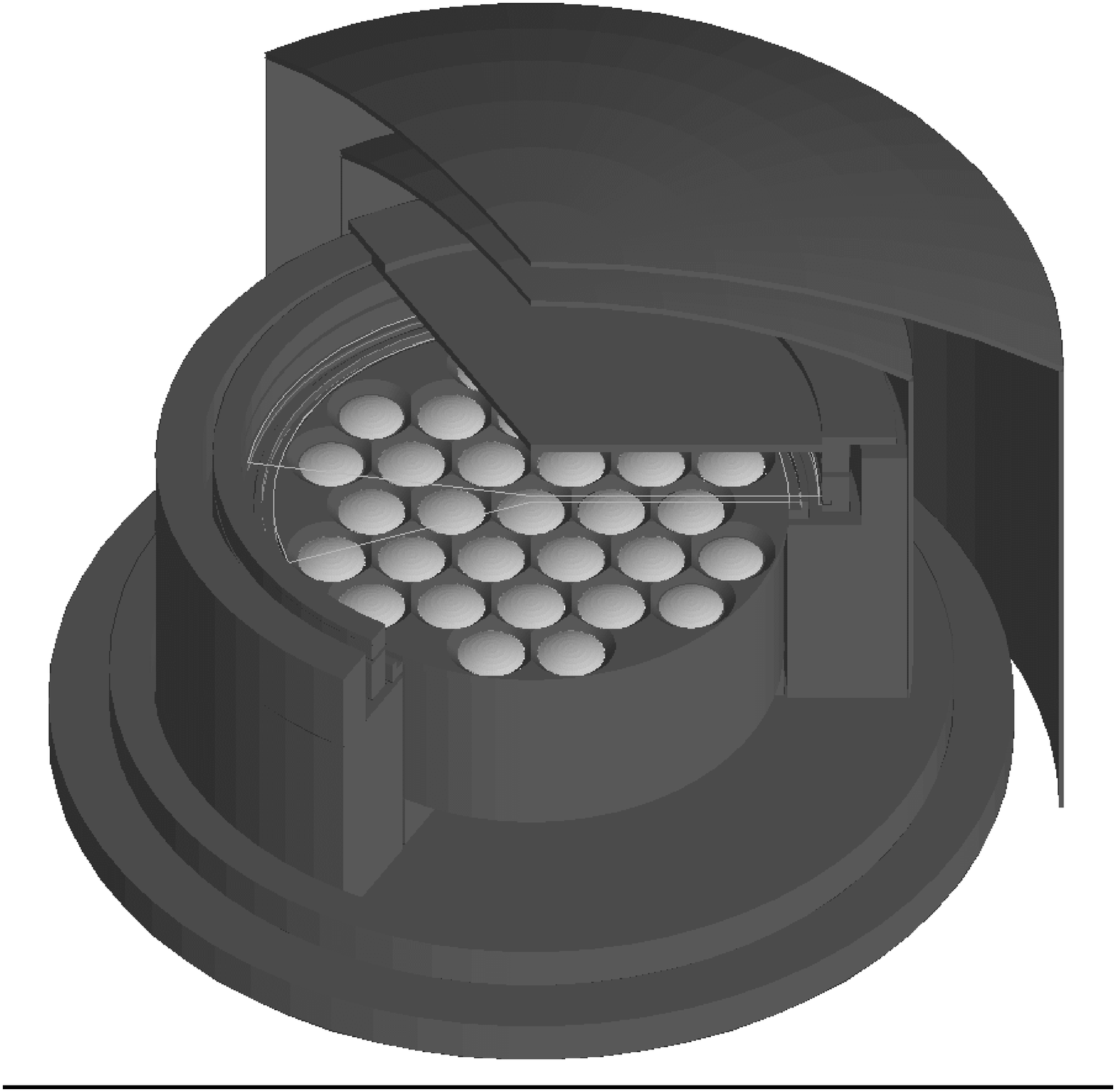,width=7.5cm,clip=} \quad
    \epsfig{file=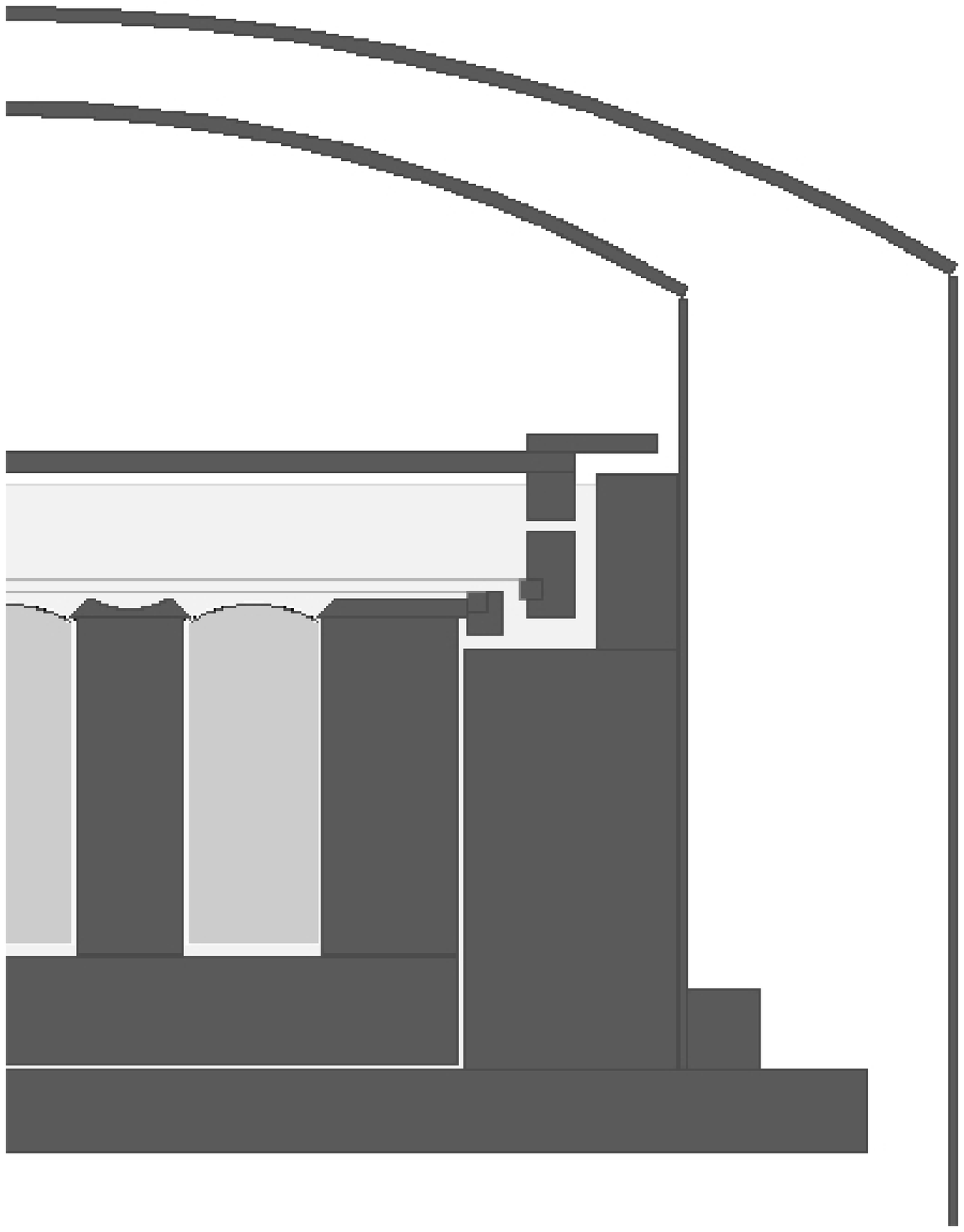,width=6cm}}
  \caption{\small Cut-out (left) and cross-sectional (right) views of
  the GEANT4 implementation of the `full' ZEPLIN-III geometry.}
  \label{geometry}
\end{figure}

Two geometry representations have been set up. The one shown in
Fig.~\ref{geometry} includes most detector components located above
the liquid nitrogen vessel. A second geometry implements only the
(optically) active region, comprised between the PMT photocathodes and
the anode mirror. Less computationally demanding than the full model,
this is useful for light collection and similar studies. Both produce
exactly the same optical response. The LXe height, chosen by default
to be 35~mm above the cathode (leaving 5~mm of gas below the anode
mirror) can also be set between runs. This entails modifying the
electric field distribution as well as optical and other properties,
as explained below. Additional external components will be included at
a future stage, namely a scintillator veto system surrounding the
detector as well as hydrocarbon and lead shielding for external
neutrons and $\gamma$-rays, respectively.

\subsection{The optical model}

Optical ray-tracing in the detector takes into account the optical
constants of liquid and gaseous xenon, the angular reflectivities of
copper surfaces and electrode grids, and the optical and detection
properties of the PMTs, all of which are defined at the xenon
scintillation wavelength of $\sim\!175$~nm (7~eV).

For LXe a refractive index $n$=1.69 and an attenuation length of
36.4~cm were considered \cite{solovov04}. Although the latter is
probably dominated by Rayleigh scattering (c.f. Ref.~\cite{seidel02},
as well as preliminary results from our ZEPLIN-II detector), only
photo-absorption was considered here -- the difference in light
collection is relatively small.

The angular reflectivity of copper is quite uncertain at VUV
wavelengths, depending on the surface finish, oxidation state and
possible LXe condensation onto the cold surfaces. A single measurement
was found, indicating $R$=25\% for normal incidence for a clean-cut
surface \cite{handbook}; a more conservative 15\% was adopted
instead. The GEANT4 `unified' model \cite{unified} was chosen to treat
the angular reflection from metal surfaces; the parameter
$\sigma_\alpha$, which characterises the Gaussian smearing of the exit
angle in this model, was set to $20^{\rm o}$, producing a half-width
$\approx$45$^{\rm o}$ for normal incidence. We point out that the
choice of model is not informed by experimental data -- lacking for
many materials for this wavelength range. A high-reflectivity case
($R$=90\% for anode and PMT mirrors) was also considered, to assess
the benefit of electroplating the mirrors in a future upgrade.

The two electrode grids (wire-wound) are implemented as continuous
dielectric sheets with refractive index matched to that of LXe (no
Fresnel reflection) but with an absorption length chosen to give 10\%
absorption at normal incidence (equal to the grid wire/pitch
ratio). The dielectric absorption mimics the obscuration of light with
angle of incidence to good approximation for all but the largest
angles with respect to the grid normal.

The PMT models consider a curved quartz window supporting an opaque
photocathode with a user-defined quantum efficiency (QE). A photon is
detected as a photoelectron depending on the outcome of a random
throw. Note that manufacture-quoted QEs are usually measured in air or
vacuum. A slightly better optical match is obtained when PMTs are
immersed in LXe due to the similar refractive indices ($n$=1.6 for
quartz). However, given that other losses are also considered
(e.g. window absorption and reflection from metal fingers deposited
onto it for improved low-temperature performance), we can assign the
measured PMT QE to the photocathode without major error.
Low-temperature QEs average 30\% for xenon scintillation for the
ZEPLIN-III phototubes \cite{araujo04}.

\subsection{Electric fields}

\begin{figure}[ht]
  \centerline{\epsfig{file=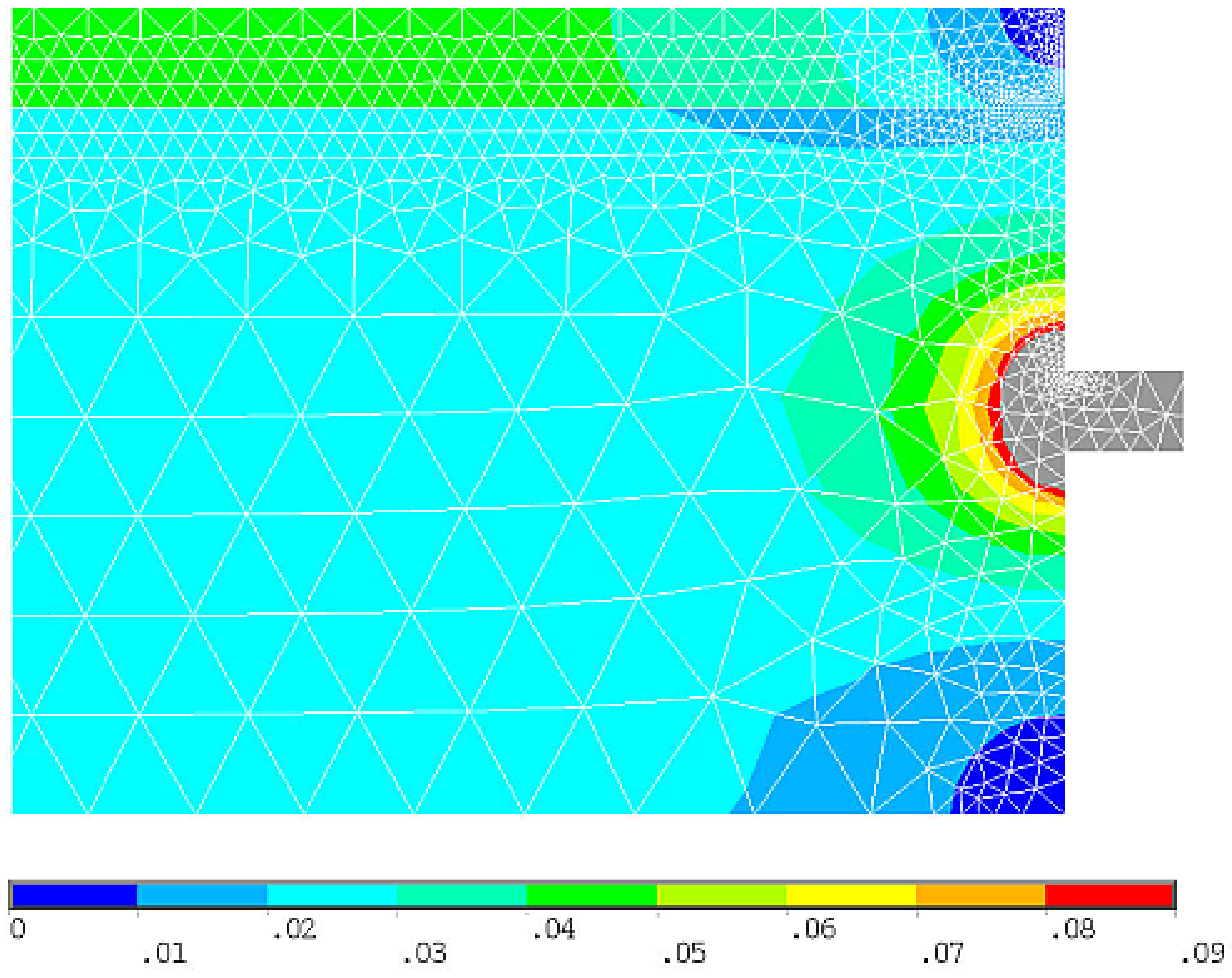,width=6.5cm,clip=} \qquad 
    \epsfig{file=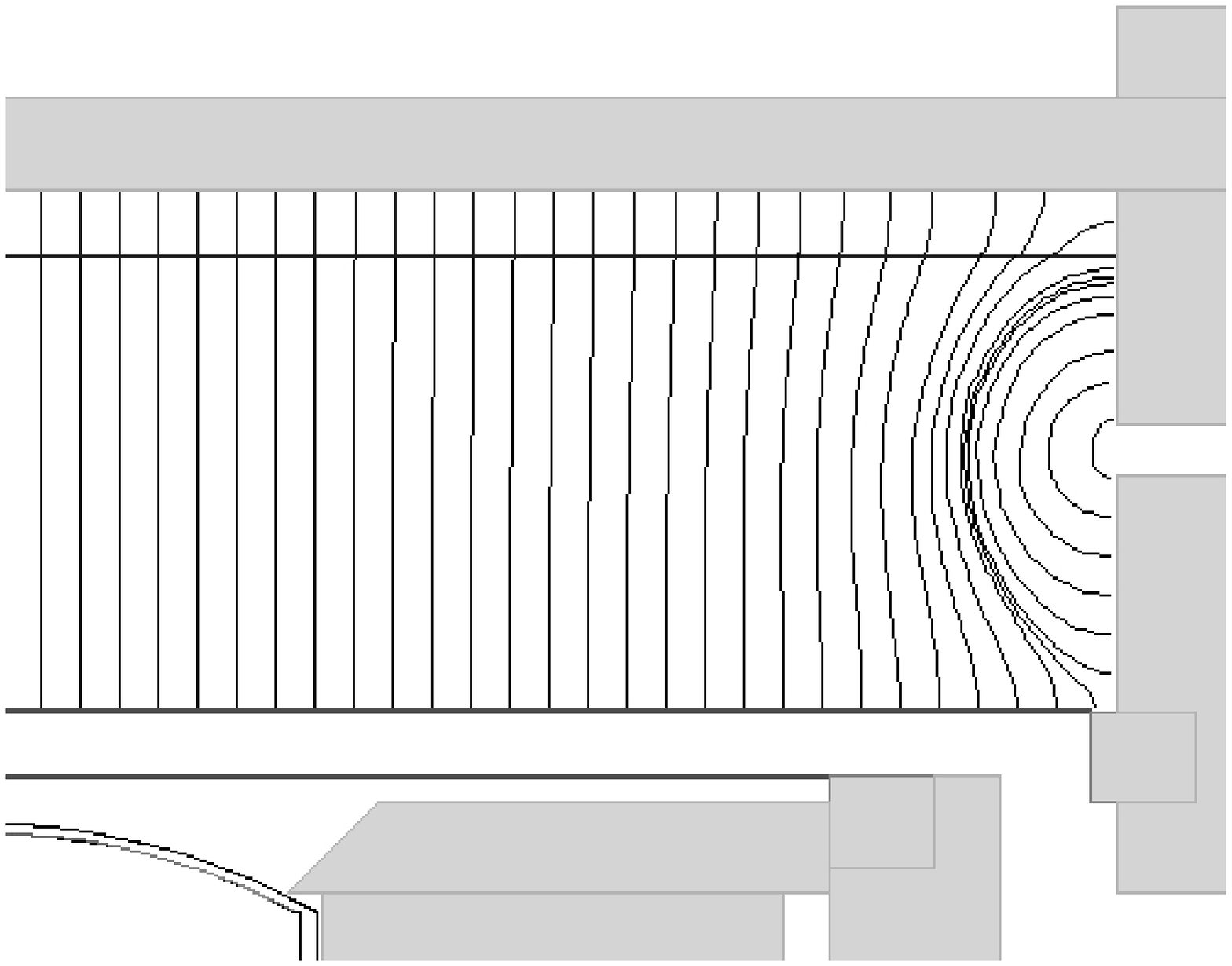,width=6.5cm,clip=}} 
  \caption{\small Left: ANSYS \cite{ansys} electric field contour map
  and optimised meshing in peripheral regions of the detector. Right:
  Simulated trajectories for charge released at the cathode grid and
  side electrode.}
  \label{field}
\end{figure}

The flat, disk-shaped LXe target ensures uniform electric fields (and
light collection) above the PMT array. Inside this central
(`fiducial') volume, high discrimination efficiency should be
achieved. Interactions in outlying regions, which can probe
non-vertical fringe fields, may still be corrected (or rejected) by
use of 3-dimensional event information. To ensure that this procedure
is well understood, the simulation uses detailed electric field models
for both phases. This is particularly important for the
electroluminescence signal: the field determines not only how much
ionisation is extracted from the interaction site, but also its drift
time to the surface, the emission probability and location, the
electroluminescence yield in the gas and the temporal development of
the signal.

Two-dimensional electric field maps are used by the
electroluminescence model; by exploiting the cylindrical symmetry of
the target, these maps can be kept small and are easily navigated. The
standard GEANT4 tracking kernel is applied to all other particles, so
there is no penalty to other processes. The ANSYS \cite{ansys}
finite-element software was used to produce field maps for several gas
gaps, for a reference voltage of 1~V between the anode mirror and the
cathode grid. The reverse-field region below the cathode grid, which
suppresses S2 signals from low-energy PMT photons, is not implemented.
Dielectric constants for the gas and liquid phases are $\epsilon_g$=1
and $\epsilon_l$=2, respectively. The total inter-electrode voltage
(set interactively) scales the chosen field map to obtain the correct
electric field strength.  Fig.~\ref{field}~(left) represents the ANSYS
contour plot and the optimised meshing for the default map (5~mm
gas). Electron trajectories obtained with a simple field navigation
are shown in Fig.~\ref{field}~(right). Note that the field is constant
above the array, well away from the edge.

\subsection{The primary signal}

\begin{figure}[ht]
  \centerline{\epsfig{file=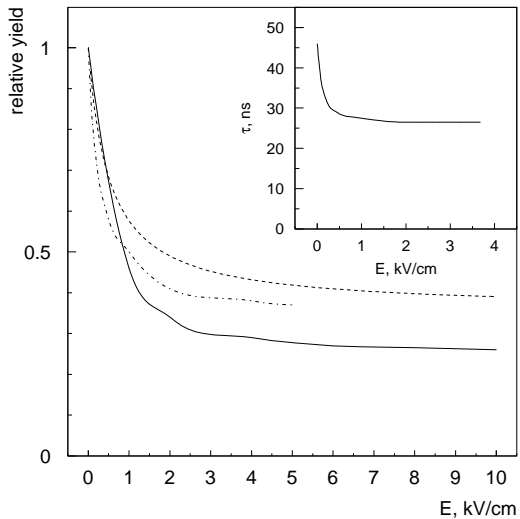,width=7.5cm}}
  \caption{\small Electric-field dependence of the scintillation yield
   for 1 MeV electrons \cite{kubota79} (continuous line),
   $\sim$200~keV Compton electrons \cite{dawson04} (dashed) and
   122~keV $\gamma$-rays \cite{aprile05} (dash-dot). The inset shows the
   decay time constant (pulse area/amplitude) for $\sim$200~keV
   Compton electrons \cite{dawson04}.}
   \label{scint}
\end{figure}

As a scintillator LXe compares favourably with the best crystals,
yielding some 60 photons per keV of energy deposited (electron
equivalent, hereafter `keVee') \cite{doke99,chepel05b}. The VUV
luminescence is produced by the decay of singlet and triplet states of
the Xe$_2^*$ excimer. These can be excited directly by the interacting
particle or as a result of recombination along the particle track
(Xe$_2^+$+$e^-$$\rightarrow$Xe$_2^*$)\cite{kubota79,hitachi83}.
Recombination is very efficient for nuclear recoil interactions, and
so their scintillation is faster ($\tau\!\simeq\!21$~ns in a
single-exponential approximation) than for electrons
($\tau\!\lesssim\!45$~ns, depending on energy) \cite{kubota79,akimov02}. This
discrimination technique was exploited in the ZEPLIN-I detector
\cite{zeplin1}.

The scintillation efficiency for nuclear recoils is lower than for
electron interactions of the same energy. A zero-field quenching
factor $QF$=0.22 \cite{akimov02} was used in the simulation regardless
of energy. Very recent data agree with this value at a few tens of keV
(nuclear-recoil energy, `keVnr'), but may indicate some reduction at
lower energies \cite{chepel05,aprile05}.

An external electric field suppresses recombination to a varying
degree, affecting mainly the light yield and decay times for electron
recoils, as depicted in Fig.~\ref{scint}. The scintillation model
(adapted from that implemented in GEANT4) assumes a uniform electric
field across both phases, i.e. yields and time constants are
position-independent. This is not unreasonable considering that, for
the range of interesting fields, these properties do not vary
appreciably. The default yields for electron and nuclear recoils were
set at 18~photons/keV and 12~photons/keV, respectively, at the nominal
8~kV/cm field. The number of VUV photons is Gaussian-distributed
(except when fewer than 10, in which case Poisson deviates are used),
with unity Fano factor -- note that for relatively small S1 the
overall energy resolution is instead dominated by photoelectron
statistics. Although pulse-shape analysis of S1 cannot provide as good
a discrimination as in a zero-field detector, this may still prove a
valuable diagnostic technique for calibration runs. For this reason,
some timing properties were implemented in the scintillation model: a
single-exponential decay is assumed, with time constants of 26~ns for
electrons and 16~ns for recoils.

It should be pointed out that the scintillation and ionisation signals
are correlated event-by-event, since recombination will contribute to
either one or the other \cite{conti03}. This correlation has not been
taken into account in this model. However, it can improve the
discrimination ability of two-phase detectors \cite{dawson03}.

\subsection{The secondary signal}
\label{s2}

Under the strong electric field, ionisation electrons are extracted
from the interaction site and drifted upward to the liquid surface;
once emitted into the gas phase they acquire enough energy to generate
many VUV photons by electroluminescence. The S2 signal is proportional
to the number of charge carriers extracted from the liquid, as well as
the electric field and path length in the gas.

\begin{figure}[ht]
\centerline{\epsfig{file=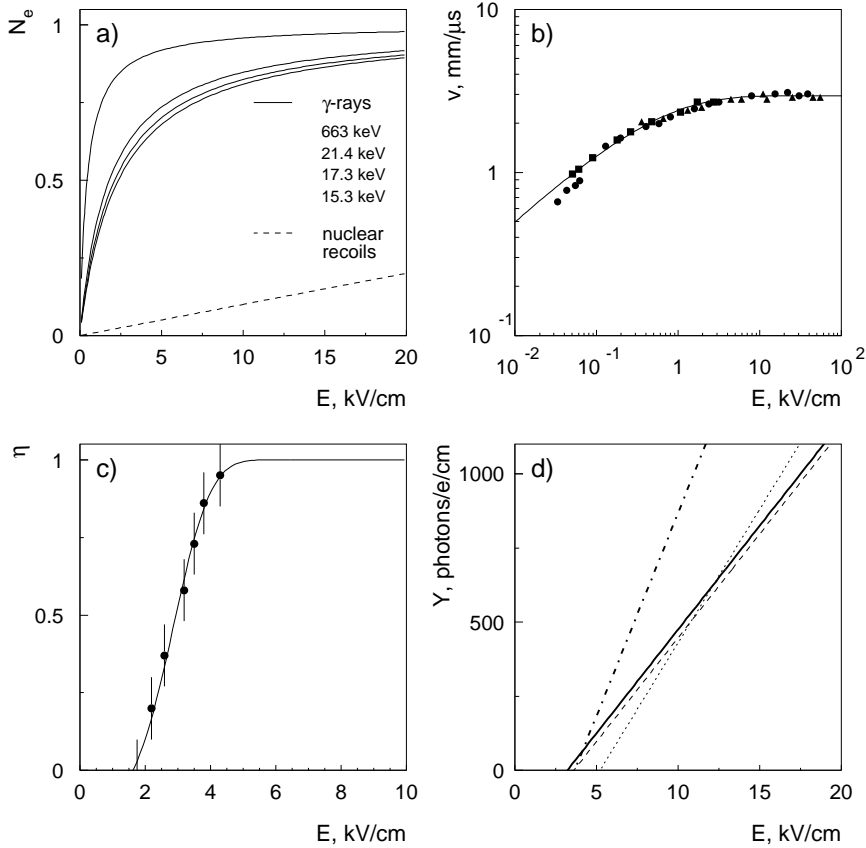,width=12cm}} 
\caption{\small Electric-field dependence of S2 parameters: a)
  Relative ionisation yield adopted for $\gamma$-rays and nuclear
  recoils in LXe; $\gamma$-ray curves are parameterised from
  experimental data from Ref.~\cite{voronova89} (ordered as in
  legend). b) Drift speed in LXe, parameterised from data found in
  Ref.~\cite{miller68}. c) Electron emission probability at liquid/gas
  interface, parameterised from experimental data in
  Ref.~\cite{gushchin79}. d) Electroluminescence yields for gaseous
  xenon at 4~bar equivalent pressure at 0$^{\rm o}$C; continuous line:
  $Y\!=\!70E\!-\!56P_{eq}$ \cite{akimov98}; dashed line:
  $Y\!=\!70E\!-\!63P_{eq}$ \cite{parsons89}; dotted line:
  $Y\!=\!90E\!-\!117P_{eq}$ \cite{ngoc79}; dash-dotted line: saturated
  vapour, $Y\!=\!137E\!-\!125P_{eq}$ \cite{fonseca04}.}
\label{elum}
\end{figure}

The number of electrons escaping recombination near the interaction
site in LXe is calculated according to the electric field strength,
the type of interacting particle and its energy. An energy $E_d$
deposited in the liquid creates $E_d/W$ free carriers, where $W$ is
the mean energy required to create an electron-ion pair at infinite
field. For $\gamma$-like interactions, we adopted $W_e$=15.6~eV
\cite{takahashi75}. Some of the ionisation produced will recombine at
finite field -- this fraction depending on the $\gamma$-ray energy.
Note that our present understanding of charge recombination is not
accurate enough to constitute a microscopic model; for this reason,
the dependence on particle energy (as opposed to energy deposited in
LXe) is the only link to experimental data. We define a fractional
charge yield, $N_e$, representing the number of carriers that escape
recombination, as:
\begin{equation}
   \frac{1}{N_e}=\left(\frac{30}{E_\gamma}+0.4\right)\frac{1}{E}+1.0 ,
\end{equation}
where $E_{\gamma}$ is the $\gamma$-ray energy in keV and $E$ is the
field strength in kV/cm \cite{davidge03}. This function parameterises
experimental data found in Ref.~\cite{voronova89} and is plotted in
Fig.~\ref{elum}~a) for the photon energies considered in that study.

The ionisation yield for nuclear recoils was unknown at the start of
this work, but it was thought to be much smaller than that for
electron recoils. It was reasoned that, to first approximation, the
number of ionisations and that of excitations producing scintillation
would be suppressed by a similar amount,
i.e. $W_n$=$W_e$/QF$\simeq$71~eV. Under a finite electric field, only
a fraction of this ionisation can be extracted from the nuclear
track. Establishing a parallel with $\alpha$-particle ionisation
yields, we adopted a linear dependence with field giving $10\%$ charge
yield at 10~kV/cm. Combining these two figures, we predicted that
ZEPLIN-III should produce some 40 electrons from a 30~keV nuclear
recoil (1.3~e$^-$/keVnr). Recent measurements suggest that
nuclear-recoil ionisation may not be a linear function of either
recoil energy or electric field, and could be 3--4 times higher than
anticipated at 30 keVnr at half the maximum ZEPLIN~III field
\cite{aprile06}. In view of this, some simulations were repeated with
a yield 4 times higher than the default value (5.2~e$^-$/keVnr), but
still proportional to recoil energy and field.

The charge carriers released from each interaction site are tracked in
the electric field until reaching either the liquid surface or a
side-electrode. The drift speed in the liquid has been parameterised
from experimental data \cite{miller68}, as shown in
Fig.~\ref{elum}~b). The high-field saturation value is
$\simeq$3~mm/$\mu$s. Neither charge trapping by impurities nor carrier
diffusion are considered at this stage in the simulation. For a
diffusivity $D$$\sim$50~cm$^2$/s \cite{doke82}, one can estimate that
the charge arrival times at the liquid surface will be smeared by
0.1~$\mu$s following a 10~$\mu$s drift. Although this effect is not
negligible when compared to the width of S2 itself, it can be taken
into account at a later stage.

Upon reaching the surface, the normal component of the electric field
in the liquid determines the fraction $\eta$ of charge emitted into
the gas phase. Experimental data \cite{gushchin79} has been
parameterised as shown in Fig.~\ref{elum}~c). The ionisation is then
tracked in the gas field, generating electroluminescence photons along
the way. A reduced drift speed of 1.5~mm/$\mu$s/(V/cm/torr) is
considered, obtained by extrapolating data in Ref.~\cite{santos94} to
high fields.

The electroluminescence yield, defined as the number of VUV photons
created per extracted carrier and per cm of track, is calculated
according to $Y$=$70E\!-\!56P_{eq}$, where $E$ is the field in kV/cm
and $P_{eq}$ is the equivalent pressure in bar for the same gas
density at 0$^{\rm o}$C \cite{akimov98}. For a vapour pressure of
2.5~bar (P$_{eq}$=4~bar) and a nominal field of 17.8~kV/cm, a single
electron extracted can produce $\sim$500~photons over a 5~mm gas
layer. This and similar yield equations, also shown in
Fig.~\ref{elum}~d), have been found for room-temperature gaseous
xenon. Recently, photon yields measured with saturated xenon vapour in
equilibrium with the liquid phase were shown to be higher than those
observed in the warm gas (the latter being consistent with previous
experiments) \cite{fonseca04}. This brings the prospect of even higher
gain in the S2 channel.

The energy resolution achievable in S2 is dominated by fluctuations of
the number of ionisation electrons extracted from the track, with
lesser contributions from the ensuing stages (emission into the gas,
electroluminescence and photon detection). It has long been
acknowledged that the intrinsic resolution achievable by charge
readout in LXe is worse than both the calculated Fano factor
($F$=0.04, \cite{doke82}) and the Poisson limit ($F$=1) -- see
e.g. \cite{conti03}. This has been attributed to the (small-number)
statistics of $\delta$-ray production acting together with charge
recombination, which can persist even under electric fields as high as
20~kV/cm \cite{thomas88}. This effect is likely to be less marked for
energies up to a few tens of keVee. Although most results presented
here are for unity Fano factor, $F$=0 and $F$=10 were also considered.
Further (Gaussian) deviates are applied at the photon generation stage
to smear the number of photons produced per electron extracted.

\subsection {DAQ Model}

\begin{figure}[ht]
  \centerline{\epsfig{file=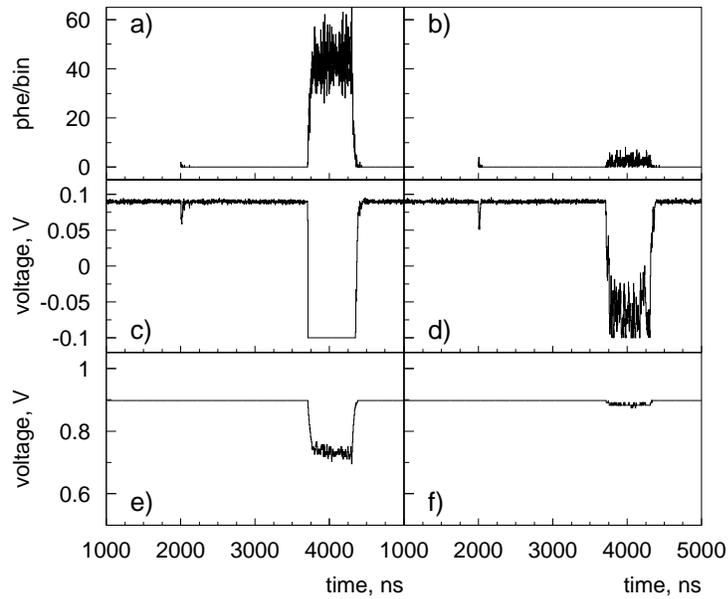,width=10cm}}
  \caption{\small Simulated timelines (31-channel sum) for 10~keVee
    electron (left) and nuclear recoil interactions (right). a) and b)
    are photoelectron timelines; c) and d) show the high-sensitivity
    DAQ channel; e) and f) show the low-sensitivity DAQ channel.}
  \label{timelines}
\end{figure}

In each event optical tracking ends with the probabilistic detection
of photons arriving at each PMT photocathode. The phe creation times
are then histogrammed into long timelines with 2~ns binning. Two such
timelines of the sum of all 31 channels are shown in
Fig.~\ref{timelines} for 10~keVee electron and nuclear recoils
(15~keVnr) interacting 5~mm below the liquid surface.\footnote{Note
that keVee is defined as the equivalent (visible) energy for electrons
taking into account the field-induced suppression of S1, so
keVee=(QF/0.3)keVnr $\simeq$2/3\,keVnr.} The two S2 signals differ by
a factor of 10 in area for a similar S1 ($\simeq$10~phe, just visible
at 2000~ns).

To obtain the PMT output voltage, the impulse response function
(i.e. single phe response) is calculated analytically \cite{wright03}
and numerically convolved with the phe timeline \cite{recipes}. For a
gain of $2\times\!10^5$, the mean single phe height is 0.16~mV and the
pulse area is 1.6~pVs. Such low gain should avoid saturation effects
following very large S2 signals. Wideband amplifiers add further gain
($\times$50) and noise (30~$\mu$Vrms at input) to the signal. Finally,
the voltage is digitised with 8-bit resolution. The dual-range DAQ
records two traces per channel at 100$\times$ full-scale
difference. Fig.~\ref{timelines} c)--f) shows how the summed timeline
would appear on both (note that noise from a single channel is
considered).

%%%%%%%%%%%%%%%%%%%%%%%%%%%%%%%%%%%%%%%%%%%%%%%%%%%%%%%%%%%%%%%%%%%%%%%%%%%%%%
\section{Simulation results}

%%%%%%%%%%%%%%%%%%%%%%%%%%%%%%%%%%%%%%%%%%%%%%%%%%%%%%%%%%%%%%%%%%%%%%%%%%%%%%
\subsection{Simulations of optical response}

\begin{figure}[ht]
\centerline{ \epsfig{file=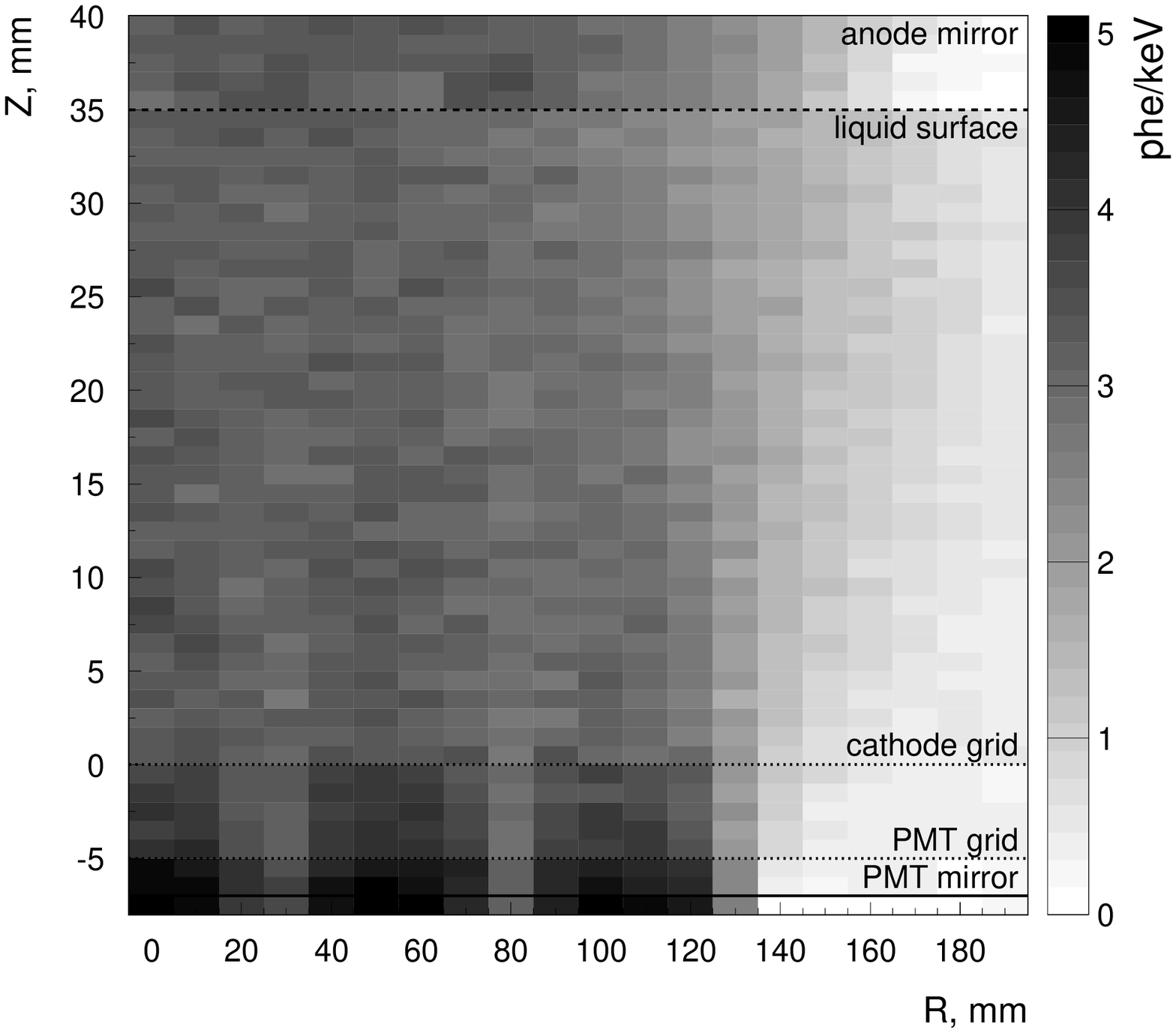,width=7.3cm}
  \epsfig{file=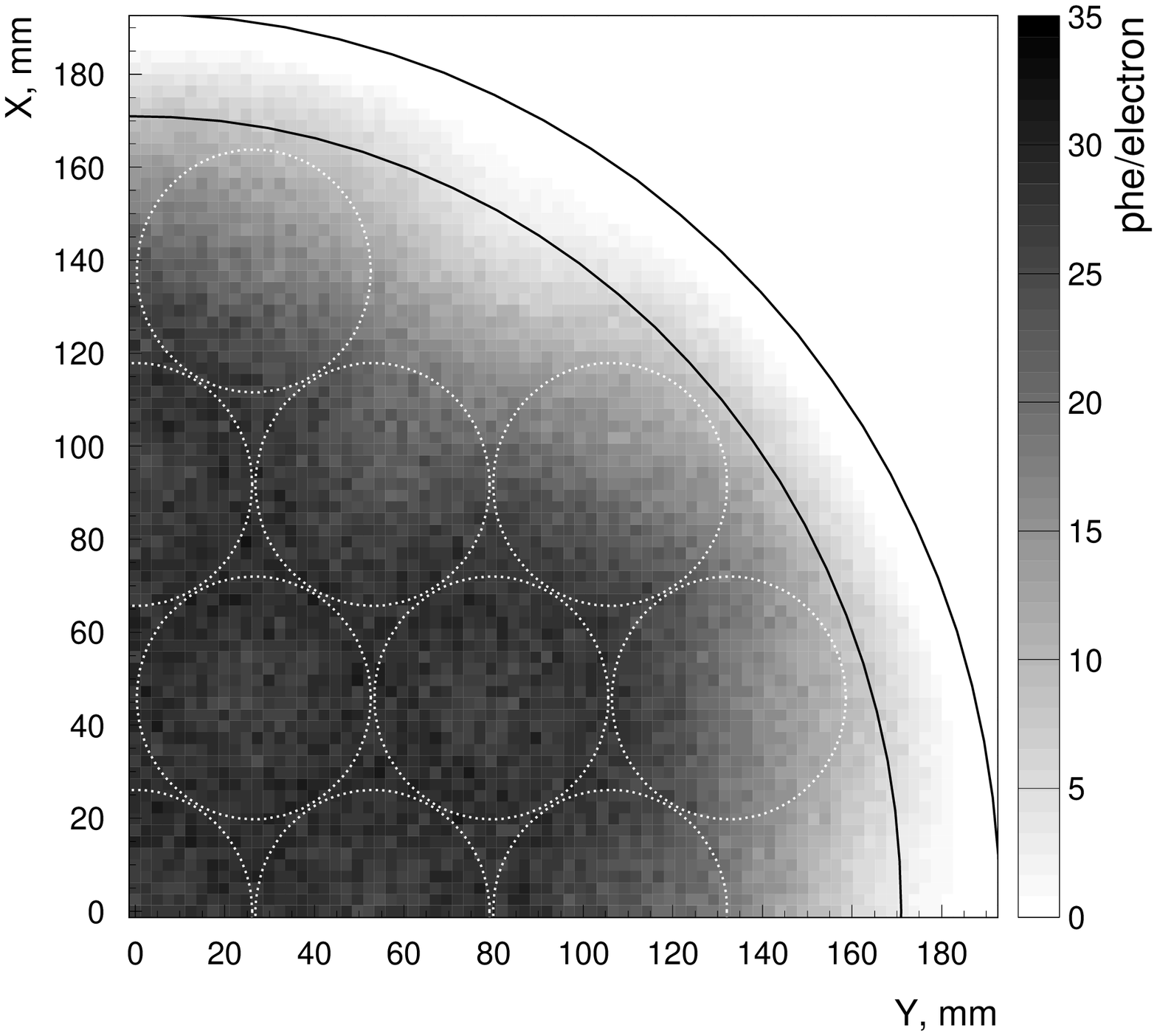,width=7.3cm}}
  \caption{\small Left: light yield from primary scintillation (S1)
  with no electric field. Right: light yield from electroluminescence
  at nominal field.}
  \label{light}
\end{figure}

The optical sensitivity to S1 and S2 are key parameters which affect
the detector performance at many levels. The primary scintillation
yield across the chamber is shown in Fig.~\ref{light} (left) for the
default parameters (5~mm of gas, 34.6~cm LXe absorption length, 15\%
copper reflectivity). It assumes a zero-field scintillation yield of
60~photons/keV in both phases. The light collection is quite uniform
across the active LXe region; a reference value for the centre of the
target is 3.4~phe/keV at zero field (1.0~phe/keV at maximum field for
electron recoils).

Increasing the mirror reflectivity to $R$=90\% would improve the
zero-field value to 4.2~phe/keV. This relatively modest increase is
due to the fact that most scintillation photons are internally
reflected by the liquid/gas interface and not by the top electrode
(and many that do reach the top plate are externally reflected at the
liquid surface and remain trapped in the gas phase). Filling the
detector to immerse the top mirror decreases the yield from
3.4~phe/keV to 2.3~phe/keV. Increasing the LXe absorption length to a
more realistic 100~cm improves the reference yield to 4.0~phe/keV with
low-$R$ mirrors.

\begin{figure}[ht]
  \centerline{\epsfig{file=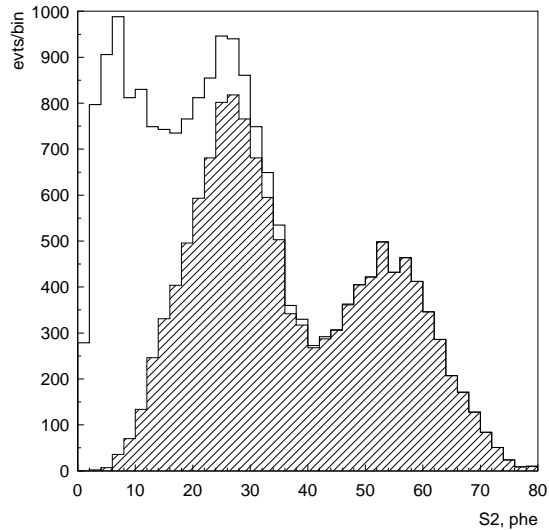,width=7.5cm}}
  \caption{\small S2 spectrum for one and two ionisation electrons
   extracted from the liquid. The shaded distribution includes only
   events in the inner 8~kg (156~mm).}
  \label{1electron}
\end{figure}

The S2 light yield was simulated by generating a constant charge over
a square grid located under the liquid surface, which is extracted to
the gas phase where it generates a varying number of photons -- thus
taking into account the position-dependent electric field. The result
is shown in Fig.~\ref{light} (right). The two concentric lines
represent the cathode grid and active volume radii. The light
collection is quite uniform across the central part of the chamber,
where an average 26~phe are generated per electron extracted.
Fig.~\ref{1electron} shows the S2 distribution for one and two
electrons emitted into the gas. Two populations can be clearly seen
when outlying events are cut ($>$156~mm radius), confirming the
sensitivity to single electrons extracted from the target.

%%%%%%%%%%%%%%%%%%%%%%%%%%%%%%%%%%%%%%%%%%%%%%%%%%%%%%%%%%%%%%%%%%%%%%%%%%%%%%
\subsection{Position reconstruction}

In ZEPLIN-III a fiducial region can be identified in the target
without physical boundaries by reconstructing the three coordinates of
the interaction. Spatial sensitivity is important in low-background
detectors since surface contamination is a source of low-energy events
(e.g. nuclear recoils from $\alpha$ decay). Interactions occurring in
regions where the electric field and the light collection are not
uniform should also be excluded.

In two-phase detectors the $z$ coordinate is obtained with sub-mm
precision by the drift time in LXe, as given by the time separation
between S1 and S2. The horizontal ($x$,$y$) position is reconstructed
using a template method; this is outlined here and described in detail
elsewhere \cite{lindote05,davidge03}. An analytical method was also
shown to be viable \cite{ditlov}.

A template was generated by full optical simulation of the S2 response
produced from 20000 locations in the gas phase, organised in a square
mesh with 2.5~mm pitch. From each point $1.5\times 10^6$ photons were
released and tracked, and the signals produced in each PMT were stored
to form a template with $20000\times 31$ entries. After simulation of
a realistic event, S2 distributions are compared with the stored
template responses, and the point with the lowest $\chi^2$ is
retrieved as the best estimate of ($x$,$y$).

An important requirement of the algorithm is that it should be able to
run on-line during data acquisition. Although the event rate expected
underground is only a few events/s, it will be higher during surface
tests and calibration runs, easily reaching the hardware limit of
100~events/s. To improve the performance of the algorithm, only a sub-set
of the template is searched in each event. This population should
represent the entire template and must therefore have homogeneous
space coverage. This was achieved by selecting the sub-set using a
2-dimensional Sobol sequence -- one of a family of so-called
`quasi-random' sequences designed to cover a given interval
homogeneously \cite{recipes}. A faster local search is then performed
in the vicinity of the selected point. A 2500-point sub-set proved to
be the limit where the resolution is not affected. This allows the
algorithm to achieve a reconstruction frequency $>$200~Hz (on a 3 GHz
CPU), which is significantly faster than the maximum DAQ rate.

\begin{figure}[ht]
  \centerline{\epsfig{file=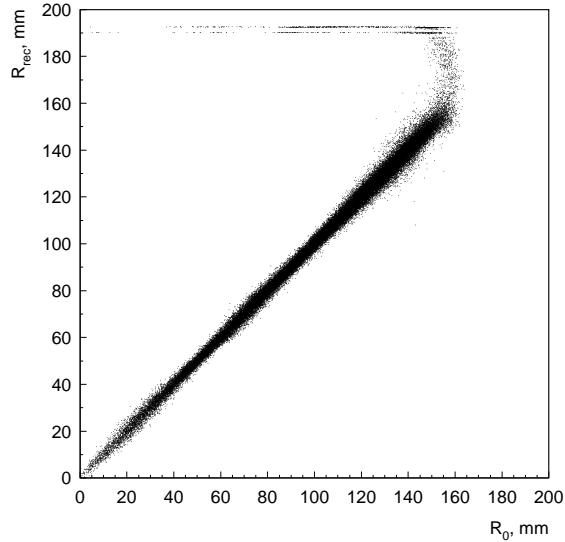,width=7.5cm}} 
  \caption{\small Plot of the reconstructed radius $R_{rec}$ as a
  function of the real radius $R_0$ of the interaction for the whole
  detector for 10 electrons extracted from the liquid.}
  \label{posrec}
\end{figure}

Random events with 1, 4 and 10 ionisation electrons extracted from the
liquid were simulated and their ($x$,$y$) coordinates reconstructed,
showing resolutions of $\simeq$10~mm (FWHM) for extremely small energy
deposits (1~electron) and a few millimetres near the threshold energy
(10~electrons). Fig.~\ref{posrec} shows the reconstructed radius for
random interactions across the whole active region for 10~electrons
extracted, subject to S2$>$10~phe (this cut eliminates fringe events
with low S2 values due to poor light collection and weaker field).

Ultimately, the merit of the algorithm must be judged by the size and
quality of the fiducial volume that can be derived. Outlying events,
with distorted signals, can be placed nearer the centre by the routine
and misinterpreted as interesting events. Contamination from these
events was studied for several test volumes, showing negligible
cross-boundary migration for up to 8~kg (156~mm radius) at very small
energy deposits (S2$<$100~phe). This is a significant improvement over
the initial design goal of 5--6~kg.

%%%%%%%%%%%%%%%%%%%%%%%%%%%%%%%%%%%%%%%%%%%%%%%%%%%%%%%%%%%%%%%%%%%%%%%%%%%%%%
\subsection{Calibration}

Calibrating the S1 and S2 channels with electron recoils of suitably
low energies throughout the detector volume is not a trivial task.
With increasing target masses, calibration with reference to a single
point or small region may prove inaccurate. S1 and S2 depend on local
electric fields and optical properties which may differ between the
LXe bulk and near electrodes and reflectors.

In ZEPLIN-III the $\gamma$ calibration strategy will consist of: i) a
relatively low-energy calibration with $^{57}$Co $\gamma$-rays for
light/charge yield measurements and long-term monitoring of the
detector stability; ii) calibration with high-energy $\gamma$-rays
from $^{60}$Co in order to populate near-threshold energies with
low-energy Compton electrons; c) low-energy $\gamma$s from inelastic
neutron scattering and from the radioactive decay of xenon and copper
radioisotopes produced by neutron activation, which can be used to
check the volume uniformity (e.g. inelastic scattering of neutrons
emitted by the PMTs cause one such feature seen at 40~keV in
Fig.~\ref{neutronbk} in the following section).

\begin{figure}[ht]
  \centerline{\epsfig{file=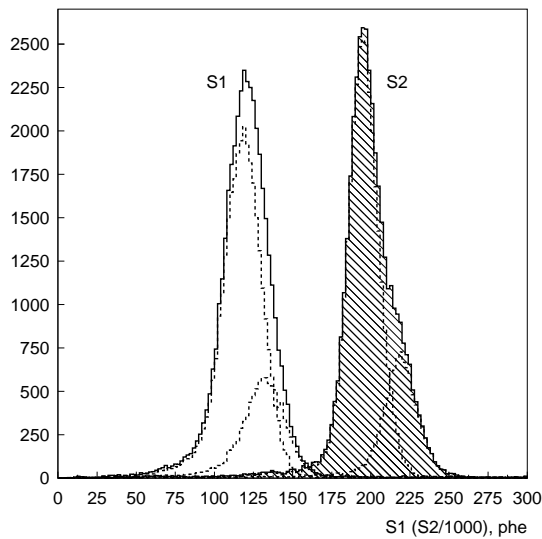,width=7.5cm}}
  \caption{\small S1 and S2 energy spectra in the inner 8~kg from
  collimated $^{57}$Co source located above the detector. S2 (shaded)
  is scaled down by a factor of 1000. The contribution of the
  individual $\gamma$ energies (122.1~keV and 136.5~keV) is also shown.}
  \label{gammacal}
\end{figure}

We outline here the simulation results for a collimated $^{57}$Co
source placed above the vacuum vessel, with reference to
Fig.~\ref{gammacal}. In spite of over 1~cm of intervening copper, a
reasonable number of 122.1~keV and 136.5 keV photons get through to
the target (14~keV $\gamma$-rays are completely
absorbed). Low-intensity lines at higher energies provide additional
calibration points, most notably the 692~keV $\gamma$-ray (not
shown). The S1 energy resolution for the 122~keV line is 25\% (FWHM),
close to the value expected from phe statistics. The energy resolution
in S2 is slightly better, being mainly determined by the Fano factor
at these energies: FWHM$\simeq\!12$\% for $F$=1. The limiting cases
are 9\% ($F$=0) and 15\% ($F$=10) -- comparable to the 11\% energy
separation between the two lines.

Calibration of the response to nuclear recoils will rely on
irradiation with an Am-Be ($\alpha$,n) source in conjunction with
independent measurements of the scintillation QF to establish the
energy scale. To calculate the recoil spectrum and the required
exposure, a 0.1~GBq Am-Be source placed above the outer vessel was
simulated. The input neutron spectrum used was that reported in
Ref.~\cite{marsh95}, with no consideration given to $\gamma$-rays from
the source. The resulting spectrum for single elastic scatters is
similar in shape to that obtained for background neutrons from the
PMTs which is shown in Fig.~\ref{neutronbk}. Some $\sim$50,000
events/hour are expected above 10~keVnr in the 8~kg fiducial volume,
with single scatters constituting approximately 60\% of all recoil
events.

Activation of xenon and other materials has also been studied and
found not to constitute a problem except for unreasonably large
exposures. Low energy signals are expected from $^{125}$I x- and
$\gamma$-rays (EC, T$_{1/2}$=59.4~days), which follows from neutron
capture on $^{124}$Xe. Other Xe isotopes and their metastable states
contribute low-energy $\gamma$s which persist for several days after
irradiation, namely $^{129}$Xe, $^{131}$Xe and $^{133}$Xe. Neutron
capture on copper is important but mainly during the exposure, since
the products are short-lived. Small amounts of $^{55}$Fe will be
created at the electrode grids, but these low-energy photons can be
cut from the data by drift-time.

%%%%%%%%%%%%%%%%%%%%%%%%%%%%%%%%%%%%%%%%%%%%%%%%%%%%%%%%%%%%%%%%%%%%%%%%%%%%%%
\subsection{Discrimination Power}

\begin{figure}[ht]
  \centerline{\epsfig{file=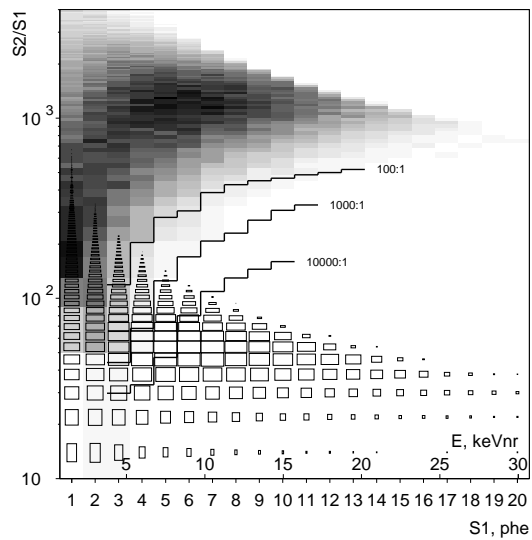,width=7.5cm}} 
  \caption{\small S2/S1 distributions for electrons (upper population)
  and nuclear recoils (lower population) in the inner 8~kg, with lower
  ionisation yield and unity Fano factor. The thick lines represent
  the boundaries for a given $\gamma$-ray discrimination
  efficiency. The energy axis considers $QF$=0.2.}
  \label{discrimination}
\end{figure}

Low-energy electrons and Xe nuclei ($5\!\times\!10^5$) were generated
randomly throughout the active volume with a constant energy
spectrum. Datasets were produced with: a) ionisation Fano factors
$F$=0, 1 and 10 and b) nuclear-recoil ionisation yields proportional
to both electric field strength and recoil energy generating an
average 12.5 and 50 electrons at 10 keVnr (the two scenarios described
in Section~\ref{s2}). The number of S1 and S2 phe in each PMT was
obtained with full optical tracking. The plot in
Fig.~\ref{discrimination} shows the ratio S2/S1 for the electron- and
nuclear-recoil populations as a function of energy ($\propto\,$S1) for
$F$=1 and the lower ionisation yield. Interactions outside the 8~kg
fiducial region were rejected.

\begin{figure}[ht]
  \centerline{\epsfig{file=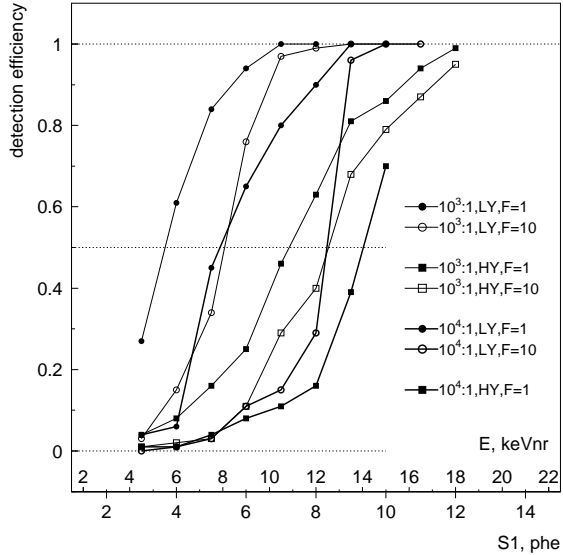,width=7.5cm}}
  \caption{\small Nuclear recoil detection efficiency curves for
  10$^3$:1 and 10$^4$:1 $\gamma$ discrimination for lower (LY, thin
  lines) and higher (HY, thick lines) ionisation yields.}
  \label{efficiency}
\end{figure}

The lines along the tail of the upper population mark $\gamma$
rejection efficiencies of 10$^n$:1 -- indicating the S2/S1 ratio at
which one event leaks to the nuclear recoil population for every
10$^n$ electron recoils. Higher rejection efficiency boundaries could
not be calculated since they would require an unreasonably long
simulation. Fig.~\ref{efficiency} plots the recoil detection
efficiencies obtained with this procedure. In conclusion, 10$^4$:1
discrimination to $\gamma$-rays can be reached between 8~keVnr and
14~keVnr for $F$=1. The combination of high ionisation yield and
$F$=10 (not shown) produces a less competitive result.

One particular class of problematic event requires consideration:
$\gamma$-rays hitting both the active region and the `dead' volume
below the cathode will produce lower S2/S1 ratios than single
interactions in the active volume, since individual S1 signals will
add up but only one S2 is generated. The discrimination efficiency
will therefore be lower for these events. Simulation of background
$\gamma$-rays emitted by the PMTs (the dominant source of electron
recoils) indicates $\sim$2700 of these multiple scatters per day. The
vast majority can be rejected by restricting the S2 location to the
fiducial volume ($<$156~mm radius) and requiring that the energy
deposits in either regions be below 10~keVee (at higher energies they
can easily be discriminated). Ten or so events per day survive these
cuts. Of these, the most dangerous ones deposit a small energy in the
target (small S2) and a comparatively large energy in the dead region
(large S1). Many such cases can be identified by analysing the S1
spatial distribution across the array: a more localised interaction
(with many phe in a single channel) is expected relative to a similar
amplitude S1 in the active region. This analysis can be extended to
take into account that the S1 location should be consistent with the
($x$,$y$) point reconstructed from S2. Very few events per day are
expected to escape these cuts, and a modest discrimination efficiency
(10$\times$ lower than that achieved in the fiducial region) should
reduce those to an insignificant contribution.

%%%%%%%%%%%%%%%%%%%%%%%%%%%%%%%%%%%%%%%%%%%%%%%%%%%%%%%%%%%%%%%%%%%%%%%%%%%%%%
\subsection{Detector Sensitivity}

The performance of ZEPLIN-III as a dark matter detector relies
ultimately on the number of background events producing signatures
which cannot be distinguished from those of WIMPs. Neutrons are the
obvious irreducible source of background, but electron recoils which
evade discrimination must also be considered. In addition, electron
recoils dominate the detector trigger rate, and this has implications
for the DAQ operation. In this section we calculate the two types of
background from dominant internal and external sources by tracking
background particles and looking for energy deposits in the
target. Full optical tracking is not performed since the very large
dynamic range of the background signals would make this computation
overwhelming.

\subsubsection{Electron-recoil backgrounds}

Trace radioactivity in the PMTs is the major source of background of
both $\gamma$-rays and neutrons -- and consequently the rate of
interactions and total energy (light) deposited in active region. The
PMT contamination levels were measured at 250~ppb in $^{238}$U,
290~ppb in $^{232}$Th and 1350~ppm in $^{40}$K. $\gamma$-rays from
$^{40}$K plus the secular equilibrium spectra for the $^{238}$U and
$^{232}$Th chains \cite{hepmcu} were generated uniformly from the PMT
glass and fully tracked until they were absorbed or left the outer
vacuum vessel (although the radioactivity is concentrated mainly at
the graded seal near the PMT windows and the glass bases, this
approximation is not unreasonable). Note that GEANT4 can generate the
$\gamma$ spectrum for the full U and Th chains, in good agreement to
the one used here, and producing in addition the correct average
number of $\alpha$, $\beta$ and $\gamma$-rays per parent decay.
Low-energy x-rays, $\alpha$ and $\beta$ radiation interacting very
close to the PMT bodies are unlikely to trigger more than one PMT due
to the copper screens in which they are located, and for this reason
were not considered.

\begin{figure}[ht]
  \centerline{\epsfig{file=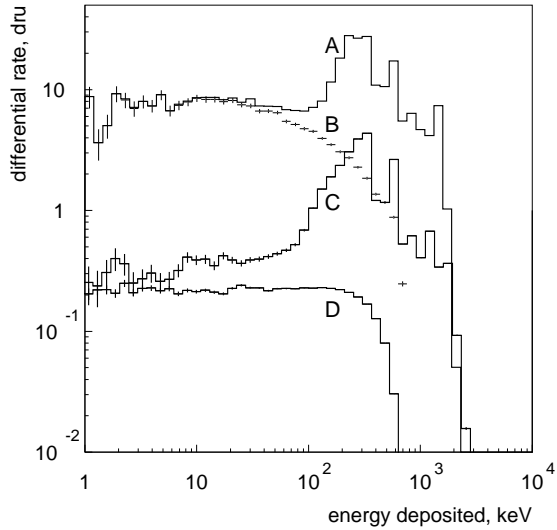,width=7.5cm}}
  \caption{\small Electron-recoil background rates. A: Total energy
  deposit in the active region from PMT $\gamma$-rays; B: Same as A,
  but for single interactions; C: $\gamma$-rays from 10~Bq/m$^3$ of
  $^{222}$Rn progeny ($^{214}$Bi+$^{214}$Pb) decaying in air inside
  the lead shielding; D) Background from $\beta^-$ decay of $^{85}$Kr
  (5 ppb Kr) in the active volume.}
  \label{gammabk}
\end{figure}

The differential energy spectrum in the target is depicted in
Fig.~\ref{gammabk} (traces A and B). We predict a low-energy rate of
10~dru (1~dru=1~evt/kg/day/keVee), confirming earlier results
\cite{davidge03}. An interaction rate of 5~Bq will be caused above the
PMT windows (S1 trigger rate), decreasing to 2.5~Bq above the cathode
grid (S2 trigger rate). This corresponds to an average rate of energy
deposited just under 2~MeV/s.

A second internal source of electron recoils is the $\beta$ decay of
$^{85}$Kr (T$_{1/2}$=10~yr) contaminating the Xe itself. Extracted
from the atmosphere in the mid 1960s, the ZEPLIN-III xenon has a low
Kr content, estimated at $\sim$5~ppb. The (Coulomb-corrected)
$\beta^-$ energy spectrum generated by GEANT4 is tracked in the
target, leading to 0.1~dru at low energies (trace~D in
Fig.~\ref{gammabk}).

The presence of $^{222}$Rn and its progeny in the air surrounding the
detector is a source of radioactive background -- mainly due
short-lived $\gamma$-emitters $^{214}$Bi and $^{214}$Pb. In
Ref.~\cite{carson05} an upper limit is placed on the differential rate
expected from these isotopes in a large Xe detector surrounded by air
contaminated with 10~Bq/m$^3$ in $^{222}$Rn; the result in reproduced
in Fig.~\ref{gammabk}~C. We note that the actual contribution from
these isotopes is likely to be even smaller, as both the air volume
and contamination level considered are too generous.

Finally, the cavern walls (mainly NaCl) emit $\gamma$-rays, requiring
extensive lead shielding around the detector. A calculation based on
radioactivity measurements of the Boulby rock (67~ppb U, 127~ppb Th,
1300~ppm K) suggests that a lead castle 15~cm thick can attenuate the
$\gamma$ background to $<$0.01~dru \cite{smith05}.

In conclusion, a total electron-recoil background of 10~dru is
expected in a shielded detector, dominated by PMT $\gamma$-rays.

\subsubsection{Nuclear-recoil backgrounds}

Energetic neutrons are produced due to radioactive contamination of
detector components and its surroundings with the uranium and thorium
decay chains, by spontaneous fission (mainly of $^{238}$U) and the
($\alpha$,n) reaction. Cosmic-ray muons also generate neutrons in
spallation reactions and secondary cascades.

Operation deep underground attenuates the muon flux by several
decades, but a few neutrons are still produced by muons interacting in
the detector materials and the cavern rock. Muon-induced neutron
production at the Boulby Underground Laboratory (2800~mwe) has been
studied in the context of a large-scale Xe detector in
Refs.~\cite{carson04,araujo05b,bungau05}. For the shielding
configuration anticipated for ZEPLIN-III one can expect $\lesssim$1
single nuclear recoil event per year above 10~keVnr in a 8~kg fiducial
volume.

The neutron flux in the cavern is instead dominated by U/Th
radioactivity in the Boulby rock. The flux at the rock face has been
simulated at $\approx\!2\times 10^{-6}$~n/s/cm$^2$ above 1~MeV
(excluding backscattering of reentering neutrons) for the above rock
contamination \cite{bungau05,carson04}. This flux would easily
overwhelm an unshielded detector. 20--30~g/cm$^2$ of hydrocarbon
shielding located inside the lead castle should attenuate the neutron
flux to acceptable levels -- producing fewer than $\sim$1 nuclear
recoil per year for a complete (4$\pi$) shield.

Neutrons produced inside the detector shielding can only be avoided by
construction with low-background materials and the use of an active
veto surrounding the detector, to record neutrons in coincidence with
the target. Construction from mainly high-purity copper (U/Th levels
$<$1~ppb) and the relatively high threshold for ($\alpha$,n) in Cu
ensure a very low neutron background from the vicinity of the target
from most components, with the notable exception of the PMTs.

\begin{figure}[ht]
  \centerline{\epsfig{file=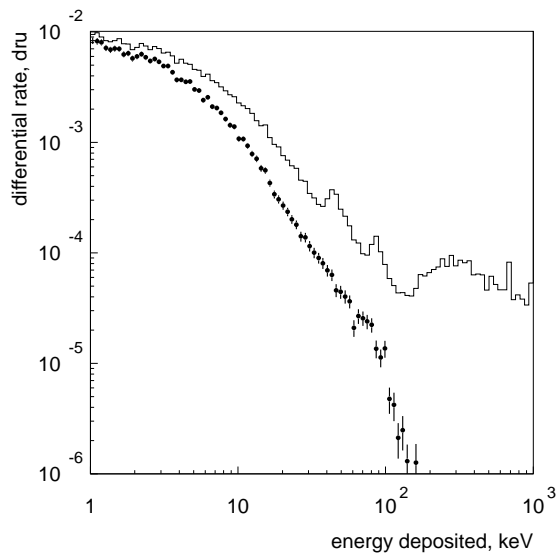,width=7.5cm}}
  \caption{\small Differential energy spectrum (evt/kg/day/keVee) from
  PMT neutrons. The line represents total energy deposited in the
  target (electron and nuclear recoils), whilst the markers show
  single nuclear recoils only (in the latter case, the abscissa
  corresponds to keVnr).}
  \label{neutronbk}
\end{figure}

Neutron spectra from fission and ($\alpha$,n) reactions in the PMT
borosilicate glass were calculated using the SOURCES-4A code
\cite{sources} modified in the way described in Ref.~\cite{carson04},
for the U/Th contamination levels mentioned
previously. Fig.~\ref{neutronbk} shows the differential energy
spectrum caused in the target by PMT neutrons emitted isotropically
and uniformly from the PMT glass. The figure represents the total
energy deposited (elastic plus inelastic events) in addition to single
nuclear-recoil events. PMT neutrons reach $\sim\!10^{-3}$~dru (e.e.)
at low energies, typically contributing 25 events/year above 10~keVnr
in the 8~kg fiducial mass.

\subsubsection{WIMP sensitivity}

Background from the 31 PMTs dominates in a shielded ZEPLIN-III by a
large margin. Some 20--40 events/year will be due to neutrons
(depending on the energy acceptance considered). In addition, although
most $\gamma$-rays can be removed by $S2/S1$ discrimination, some will
`leak' into the nuclear recoil population for any realistic
discrimination cut adopted. Increasing the latter (e.g. from 10$^3$:1
to 10$^4$:1) also raises the energy threshold, as illustrated in
Figs.~\ref{efficiency} (fortunately, it also decreases the number of
neutron events).

The yearly neutron and $\gamma$ rates are presented in Table~1 for the
ionisation yields mentioned previously (1.3 and 5.2 $e^-$/keVnr) and
$F$=1. We adopt a 50\% energy threshold, $E_0$, from the efficiency
curves in Fig.~\ref{efficiency}. The neutron distribution
(Fig.\ref{neutronbk}) is integrated above $E_0$. The number of
$\gamma$-rays is obtained by integrating a constant rate of 10~dru
(Fig.~\ref{gammabk}) over 2~keVee (very few non-discriminated
$\gamma$s should remain above $E_0$+2~keVee). The sensitivity limits
for the WIMP-nucleon cross-section (spin-independent interaction) are
also shown for the minimum of the sensitivity curve (60~GeV WIMP
mass), for an exposure of 240 kg$\times$days (1~month at 100\% duty
cycle) and 3000~kg$\times$days (1~year). These were calculated as
described in Ref.~\cite{alner05}.

\begin{sidewaystable}
\centering
\begin{minipage}{20cm}
\centering
\caption{ZEPLIN III background rates and WIMP sensitivity}
\label{table1}
\vspace{2mm}
\begin{tabular}{l|c|cc|cc|c|cc|cc}
\hline
  &\multicolumn{5}{c|}{low recoil ionisation yield}  
  &\multicolumn{5}{c}{high recoil ionisation yield} \\
  \hline
  & $E_{0}$ 
  &\multicolumn{2}{c|}{evt/yr} 
  &\multicolumn{2}{c|}{$\sigma_{min}$, pb}
  & $E_{0}$
  &\multicolumn{2}{c|}{evt/yr} 
  &\multicolumn{2}{c}{$\sigma_{min}$, pb} \\
  & keVnr & n & $\gamma$ & 240 kg$\times$days & 3000 kg$\times$days 
  & keVnr & n & $\gamma$ & 240 kg$\times$days & 3000 kg$\times$days \\
\hline
10$^3$:1 $\gamma$ disc. & 5.5& 40 & 60 &$1.0\times 10^{-7}$ &$6.7\times 10^{-8}$ 
                        & 11 & 24 & 60 &$1.5\times 10^{-7}$ &$9.5\times 10^{-8}$\\
10$^4$:1 $\gamma$ disc. & 8  & 31 & 6  &$6.9\times 10^{-8}$ &\bf{$3.4\times 10^{-8}$} 
                        & 14 & 16 & 6  &$8.8\times 10^{-8}$ &\bf{$3.6\times 10^{-8}$}\\
\hline
Veto installed\dag      & 8  & 16 & 2  &$4.7\times 10^{-8}$ &$1.8\times 10^{-8}$ 
                        & 14 & 8  & 2  &$6.3\times 10^{-8}$ &$2.1\times 10^{-8}$\\
PMT upgrade \dag\ddag   & 8  & 1.6 &0.2&$2.2\times 10^{-8}$ &$4.2\times 10^{-9}$ 
                        & 14 & 0.8 &0.2&$3.8\times 10^{-8}$ &$5.6\times 10^{-9}$\\
\hline
\end{tabular}
\footnotetext{\dag\, For $10^4$:1 $\gamma$ discrimination efficiency}
\footnotetext{\ddag\, Including veto}
\end{minipage}
\end{sidewaystable}

A limiting sensitivity of $3-4\times 10^{-8}$~pb can be achieved even
in a high ionisation yield scenario. This is due to the fact that the
reduced energy acceptance for nuclear recoils affects the neutron
background more than the WIMP-induced spectrum at these energies. This
sensitivity is similar to the original design prediction, and confirms
that ZEPLIN-III can probe deep into the parameter space favoured by
SUSY in its original design.

An active veto system installed around the detector can be used to
reject events in coincidence with the target. A veto efficiency of
50\% can be realistically achieved for internal neutrons. A
$\gamma$-ray veto efficiency of $\simeq$70\% has been deemed possible
\cite{davidge03}. A relatively modest factor of 2 in WIMP sensitivity
can be gained in this way, assuming the veto itself does not
contribute to neutron and $\gamma$ backgrounds.

A very worthwhile planned upgrade is the replacement of the phototubes
with newly-developed low-radioactivity ones. A ten-fold reduction in
background (neutrons and $\gamma$s) can be realistically
achieved. This would make PMT neutron rates in the target comparable
to external neutron backgrounds, pushing the WIMP-nucleon
cross-section sensitivity down to $\sim\!5\times 10^{-9}$~pb. With
these two upgrades in place ZEPLIN-III would compete favourably with
much larger targets and more expensive technologies being considered
around the world.

%%%%%%%%%%%%%%%%%%%%%%%%%%%%%%%%%%%%%%%%%%%%%%%%%%%%%%%%%%%%%%%%%%%%%%%%%%%%%%
\section{Conclusion}

The ZEPLIN-III performance as a WIMP detector has been assessed using
a fully-featured, realistic simulation tool based on GEANT4. The
original sensitivity of a few times $10^{-8}$~pb is confirmed and many
aspects of the detector performance have been predicted in
anticipation of tests now being carried out in the laboratory.

Some caution must be exercised in interpreting some of these results,
notably the discrimination efficiencies and the sensitivity limits
derived from them. It is hard to conceive a simulation model which
could accurately characterise the tails of the $\gamma$ and recoil
distributions to better than 1 part in 10$^4$. However, most
parameters were conservatively chosen, and the real performance could
actually surpass that predicted in this work. We shall now discuss
some of these uncertainties.

The reference light yield is likely to be higher than 3.4~phe/keV:
early results from our ZEPLIN-II detector (now operating underground)
as well as dedicated tests using ZEPLIN-III during its first
commissioning run at Imperial College \cite{chepel06} suggest that
both the scintillation yield and the photon absorption length in LXe
are higher than considered in this simulation, perhaps due to the
outstanding purity levels required for charge drift in these large
chambers. This will translate directly into an improvement in energy
threshold.

New experimental evidence \cite{aprile06} is pointing to larger
ionisation yields from nuclear recoils than originally anticipated,
more in line with the high-yield scenario considered here.
Significantly, it appears that the WIMP sensitivity is remarkably
unaffected, at least in this energy range.

The ionisation Fano factor can affect more the potential of LXe as a
WIMP target, and $F$ is not really known for electron and nuclear
recoil energies of interest. A combination of a high ionisation yield
($\sim$5~e$^-$/keVnr) and large Fano factor ($F$$\sim$10) could make
two-phase xenon detectors less attractive a technology for WIMP
searches, but this combination is not likely. In addition, we note
that the anti-correlation between S1 and S2, due to the fact that
ionisation electrons contribute to either one or the other channel,
can be exploited to improve the discrimination efficiency during data
analysis.

The new recoil ionisation data also reveal a weak dependence on
electric field above $\sim$1~kV/cm (unlike that indicated in
Fig.~\ref{elum}~a). Given that charge extraction from electron recoils
is still increasing at that field strength, the advantage of
high-field operation may prove very significant. Moreover, it is
possible that a higher field may decrease the Fano factor and lead to
better discrimination, simply because more charge is extracted from
the interaction site.

As it stands, ZEPLIN-III should be able to produce a world-beating
sensitivity before rival systems and other technologies. With a
further upgrade of the phototubes a very significant gain in
sensitivity can be achieved, since these dominate both the neutron and
$\gamma$-ray backgrounds.

%%%%%%%%%%%%%%%%%%%%%%%%%%%%%%%%%%%%%%%%%%%%%%%%%%%%%%%%%%%%%%%%%%%%%%%%%%%%%%
\section{Acknowledgements}

This work has been funded by the UK Particle Physics \& Astronomy
Research Council (PPARC).
%One of the authors (AL) acknowledges funding from the Portuguese
%{\it Funda\c{c}\~{a}o para a Ci\^{e}ncia e a Tecnologia}.
Thanks are also due the GEANT4 Collaboration for excellent user
support.

%%%%%%%%%%%%%%%%%%%%%%%%%%%%%%%%%%%%%%%%%%%%%%%%%%%%%%%%%%%%%%%%%%%%%%%%%%%%%%
%%%%%%%%%%%%%%%%%%%%%%%%%%%%%%%%%%%%%%%%%%%%%%%%%%%%%%%%%%%%%%%%%%%%%%%%%%%%%%

\end{document}